\global\def\draftcontrol{0}

   \def\versionno{ n2  order}

\catcode`\@=11

\expandafter\ifx\csname draftcontrol\endcsname\relax\global\def\draftcontrol{0}
\fi

{\count255=\time\divide\count255 by 60
\xdef\hourmin{\number\count255}
\multiply\count255 by-60\advance\count255 by\time
\xdef\hourmin{\hourmin:\ifnum\count255<10 0\fi\the\count255}}
\def\draftdate{\number\month/\number\day/\number\year\ \ \ \hourmin }

\newcommand\makepapertitle{\par
  \begingroup
    \renewcommand\thefootnote{\@fnsymbol\c@footnote}%
    \def\@makefnmark{\rlap{\@textsuperscript{\normalfont\@thefnmark}}}%
    \long\def\@makefntext##1{\parindent 1em\noindent
            \hb@xt@1.8em{%
                \hss\@textsuperscript{\normalfont\@thefnmark}}##1}%
     \newpage
     \global\@topnum\z@   
     \@makepapertitle
     \thispagestyle{empty}\@thanks
  \endgroup
  \setcounter{footnote}{0}%
  \global\let\thanks\relax
  \global\let\makepapertitle\relax
  \global\let\@makepapertitle\relax
  \global\let\@thanks\@empty
  \global\let\@author\@empty
  \global\let\@date\@empty
  \global\let\@title\@empty
  \global\let\title\relax
  \global\let\author\relax
  \global\let\date\relax
  \global\let\and\relax
  \def\version{\let\version\@version\@gobble}
}
\def\@makepapertitle{%
  \newpage
   \ifnum\draftcontrol=1 {}
   \version\versionno
   \vskip 3em%
   \else
   \hfill\hbox to 3cm {\parbox{4cm}{\@pubnum}\hss}%
   \vskip 3em%
   \fi
   \begin{center}%
   \let \footnote \thanks
     {\LARGE {\@title}}%
     \vskip 1.5em%
     {\normalsize
       \lineskip .5em%
       \begin{tabular}[t]{c}%
         \@author
       \end{tabular}\par}%
     \vskip 1.5em%
     {\@bstract}%
     \end{center}%
     \vskip 1.5em
     \@date%
   \par
}

\gdef\@pubnum{}
\def\pubnum#1{%
  \gdef\@pubnum{#1}}

\gdef\@bstract{}
\def\Abstract#1{%
  \gdef\@bstract{%
   \parbox{\textwidth-0pc}{%
   \centerline{\bf Abstract}\penalty1000%
\kern.2cm%
\noindent
\renewcommand\baselinestretch{1.0}%
{#1}}}
}

\def\ps@paper{\let\@mkboth\@gobbletwo%
     \ifnum\draftcontrol=1
    \def\@oddfoot{\hbox to \textwidth{\tiny \versionno \hfil\tiny\draftdate}%
    \hskip -\textwidth \hbox to \textwidth{\hfil\rm\thepage\hfil}}%
     \else\def\@oddfoot{\hbox to \textwidth{\hfil\rm\thepage\hfil}}
     \fi
     \let\@evenfoot\@oddfoot
}

\def\body{\clearpage
          \pagestyle{paper}
    }

\def\@version#1{\ifnum\draftcontrol=1
\typeout{}\typeout{#1}\typeout{}
\vskip3mm\centerline{\hbox{\fbox{\normalsize{\tt DRAFT -- #1 -- }
                   {\draftdate}}}}\vskip3mm
\fi}
\let\version\@version
\long\def\eqlabel#1{\ifnum\draftcontrol=1
                    \tag@false  
                    \tag*{(\theequation) \hbox to -0.2cm{\hspace{0cm}\small{#1}\hss}}
                    \refstepcounter{equation}
                    \edef\@currentlabel{\theequation}
                    \ltx@label{#1}          
                    \else
                    \label{#1}
                    \fi
                    }
\let\st@bibitem\@bibitem
\let\st@lbibitem\@lbibitem
\ifnum\draftcontrol=1
  \def\@bibitem#1{%
    \st@bibitem{#1}\a@@label{#1}\ignorespaces}
  \def\@lbibitem[#1]#2{%
    \st@lbibitem[#1]{#2}\a@@label{#2}\ignorespaces}
  \def\a@@label#1{%
    \gdef\a@lab{\smash{\normalfont\small#1}}
    \ifvmode
      \if@inlabel
        \global\setbox\@labels\hbox{%
          \llap{\a@lab\let\a@lab\relax
                \kern\@totalleftmargin\kern\marginparsep}%
          \box\@labels}%
      \fi
    \fi}
\fi

\documentclass[12pt,letterpaper]{article}

\usepackage{amsmath,amssymb,array,calc,epsfig,rotating,bm,xcolor}
\usepackage[sort]{cite}
\usepackage{graphicx,esint,float}
\usepackage{psfrag,verbatim}
\usepackage[makeroom]{cancel}
\usepackage{xcolor,url}
\usepackage{hyperref}

\ifnum\draftcontrol=1
\fi

\renewcommand\baselinestretch{1.25}
\renewcommand\section{\@startsection {section}{1}{\z@}%
                                   {-3.5ex \@plus -1ex \@minus -.2ex}%
                                   {2.3ex \@plus.2ex}%
                                   {\normalfont\large\bfseries}}
\renewcommand\subsection{\@startsection{subsection}{2}{\z@}%
                                   {-3.25ex\@plus -1ex \@minus -.2ex}%
                                   {1.5ex \@plus .2ex}%
                                   {\normalfont\normalsize\bfseries}}
\renewcommand\subsubsection{\@startsection{subsubsection}{3}{\z@}%
                                   {-3.25ex\@plus -1ex \@minus -.2ex}%
                                   {1.5ex \@plus .2ex}%
                                   {\normalfont\normalsize\it}}
\renewcommand\paragraph{\@startsection{paragraph}{4}{\z@}%
                                   {-3.25ex\@plus -1ex \@minus -.2ex}%
                                   {1.5ex \@plus .2ex}%
                                   {\normalfont\normalsize\bf}}

\numberwithin{equation}{section}

\def\revise#1       {\raisebox{-0em}{\rule{3pt}{1em}}%
                     \marginpar{\raisebox{.5em}{\vrule width3pt\
                     \vrule width0pt height 0pt depth0.5em
                     \hbox to 0cm{\hspace{0cm}{%
                     \parbox[t]{4em}{\raggedright\footnotesize{#1}}}\hss}}}}

\newcommand\nxt[1]  {\\\fnxt#1}
\newcommand{\ie}{{\it i.e.,}\ }

\newcommand{\mathcolorbox}[2]{\colorbox{#1}{$\displaystyle #2$}}

\def\calc         {{\cal C}}
\def\cald         {{\cal D}}
\def\cale         {{\cal E}}
\def\calf         {{\cal F}}

\def\calh         {{\cal H}}
\def\cali         {{\cal I}}
\def\calj         {{\cal J}}

\def\call         {{\cal L}}
\def\calm         {{\cal M}}
\def\caln         {{\cal N}}
\def\calo         {{\cal O}}

\def\calv         {{\cal V}}

\def\calx         {{\cal X}}

\def\reals        {{\mathbb R}}
\def\zet          {{\mathbb Z}}

\def\del          {\partial}

\def\Re           {{\rm Re\hskip0.1em}}
\def\Im           {{\rm Im\hskip0.1em}}

\def\sqr#1#2{{\vcenter{\vbox{\hrule height.#2pt
 \hbox{\vrule width.#2pt height#1pt \kern#1pt
 \vrule width.#2pt}\hrule height.#2pt}}}}
\def\square{%
  \mathop{\mathchoice{\sqr{12}{15}}{\sqr{9}{12}}{\sqr{6.3}{9}}{\sqr{4.5}{9}}}}

\newcommand{\ww}{\mathfrak{w}}

\def\aa1{\phi}
\def\cc1{\psi}

\def\f0{\text{\boldmath$\varphi$}}
\def\h2{\mathfrak{h}}

\catcode`\@=12

\begin{document}


\title{\bf Compactified holographic conformal order}

\date{July 11, 2021}

\author{
Alex Buchel\\[0.4cm]
\it $ $Department of Physics and Astronomy\\ 
\it University of Western Ontario\\
\it London, Ontario N6A 5B7, Canada\\
\it $ $Perimeter Institute for Theoretical Physics\\
\it Waterloo, Ontario N2J 2W9, Canada
}

\Abstract{
We study holographic conformal order compactified on $S^3$.  The
corresponding boundary CFT$ _4$ has a thermal phase with a nonzero
expectation value of a certain operator. The gravitational dual to the
ordered phase is represented by a black hole in asymptotically $AdS_5$
that violates the no-hair theorem. While the compactification does not
destroy the ordered phase, it does not cure its perturbative
instability: we identify the scalar channel QNM of the hairy black
hole with $\Im[\ww]>0$. On the contrary, we argue that the disordered
thermal phase of the boundary CFT is perturbatively stable in
holographic models of Einstein gravity and scalars.
}

\makepapertitle

\body

\version\versionno
\tableofcontents

\section{Introduction}\label{intro}
{\it Conformal order} stands for exotic thermal states of conformal theories
with spontaneously broken  global symmetry group $G$
\cite{Chai:2020zgq,Chaudhuri:2020xxb,Chai:2021djc,Chaudhuri:2021dsq}.
For a CFT$ _{d+1}$ in Minkowski space-time $\reals^{d,1}$ the existence of the
ordered phase implies that there are at least two distinct thermal
phases:
\begin{equation}
\frac{\calf}{T^{d+1}}=-\calc\ \times\
\begin{cases}
1,\qquad T^{-\Delta}\langle\calo_\Delta\rangle=0\ \Longrightarrow\
G\ {\rm is\ unbroken}; \\
\kappa,\qquad T^{-\Delta}\langle\calo_\Delta\rangle=\gamma\ne 0\ \Longrightarrow\
G\ {\rm is\ spontanuously\ broken},
\end{cases}
\eqlabel{phd}
\end{equation}
where $\calf$ is the free energy density, $T$ is the temperature,
$\calc$ is a positive constant
proportional to the central charge of the theory, and $\calo_\Delta$ is the
order parameter for the symmetry breaking of conformal dimension
$\Delta$. The parameters $\kappa$ and $\gamma$
characterizing the thermodynamics of the
the symmetry broken phase are  necessarily constants. Note that
when $\kappa>1$ ($\kappa<1$), the symmetry broken phase dominates
(is subdominant) both
in the canonical and the microcanonical ensembles.
Irrespectively
of the value, provided $\kappa>0$, the symmetry broken phase is
thermodynamically stable. It is difficult to compute directly in
a CFT the values $\{\kappa, \gamma\}$, thus establishing 
the present and the (in)stability of the ordered phase. 
Rather, the authors of
\cite{Chai:2020zgq,Chaudhuri:2020xxb,Chai:2021djc,Chaudhuri:2021dsq}
established the instability of the {\it disordered} (symmetry preserving)
thermal phases in discussed CFTs. The condensation of the
identified unstable mode then leads to $\langle\calo_\Delta\rangle\ne 0$
for the new equilibrium thermal state --- the conformal order. 

Conformal order states are very interesting in the context of
holography \cite{Maldacena:1997re,Aharony:1999ti},
as they imply the  existence of the dual black branes in a Poincare patch of 
asymptotically $AdS_{d+2}$ bulk geometry that violate the no-hair theorem.
In section \ref{stable} we prove a theorem that the disordered
conformal thermal states are always stable\footnote{We consider $AdS_5/CFT_4$
dualities, but the argument can be readily extended to other dimensions.
} in dual holographic models
of Einstein gravity with multiple scalars.
Thus, the mechanism for the conformal order presented
in \cite{Chai:2020zgq} is not viable in these holographic models.

The no-go theorem of section \ref{stable} does not imply that
the holographic conformal order can not exist. In fact, way before
\cite{Chai:2020zgq}, what is now known as a conformal order was constructed
in \cite{Buchel:2009ge}\footnote{See \cite{Buchel:2020thm,Buchel:2020xdk}
for the recent work. Potentially relevant top-down constructions
also include \cite{Buchel:2018bzp,Buchel:2021yay}.}. In the gravitational
dual to the holographic conformal order it is straightforward to
compute the parameter $\kappa$ in \eqref{phd}, and establish that
$\kappa<1$: at best, the holographic conformal order is
metastable\footnote{Given the no-go theorem reported here this is
not surprising.}.  In \cite{Buchel:2020jfs} it was established
that the holographic conformal order is perturbatively unstable.
Specifically, we identified a non-hydrodynamic quasinormal mode (QNM)
of the dual hairy black brane with\footnote{We define
dimensionless frequencies as $\ww\equiv w/(2\pi T)$.} $\Im[\ww]>0$.
The purpose of this paper is to use the techniques of
\cite{Buchel:2021ttt} to investigate the stability of the holographic
conformal order compactified on $S^3$.

As introduced, the conformal order
is associated with the spontaneous breaking of a global symmetry.
Thus, one might wonder whether the order can ever survive the
compactification. To this end, we point the following:
\nxt In the infinitely large-$N$ (large central charge)
limit spontaneous  symmetry
breaking can  happen on a compact manifold. For a recent
example, see a discussion of the spontaneous chiral symmetry breaking of the
cascading gauge theory on $S^3$ \cite{Buchel:2021yay}. 
\nxt We present an explicit example in this paper where the existence of the
holographic conformal order is not associated with the breaking of any
global symmetry.

A detailed summary of our study is provided in section \ref{sum}.
We highlight here the new results:
\begin{itemize}
\item Our main model --- $\mathfrak{M}_{PW}^b$ --- is a  holographic $CFT_4$ with a single
operator $\calo_2$ of the conformal dimension $\Delta=2$. The order is
associated with the thermal expectation value $\langle\calo_2\rangle\ne 0$.
The subscript $ _{PW}$ indicates that the model is a deformation,
$b$ is the deformation parameter, of the top-down $\caln=2^*$
gauge theory/gravity correspondence
\cite{Pilch:2000ue,Buchel:2000cn,Buchel:2013id,Bobev:2013cja}.
$\caln=2^*$ holographic correspondence is recovered in the limit $b\to 1$.
Importantly, our conformal model $\mathfrak{M}_{PW}^b$ does not have
any global symmetry. 
We follow the techniques introduced in \cite{Buchel:2020xdk} and construct
conformal order in $\mathfrak{M}_{PW}^b$, with the holographic
boundary CFT in $\reals^{3,1}$ Minkowski space-time, perturbatively
as $b\to \infty$. This is the first ever example of the holographic
conformal order
that is not associated with spontaneous breaking of a global symmetry.
We find that $\kappa<1$ (see \eqref{phd}) in the model, making the ordered
phase not the preferred one. We identify the unstable QNM mode of the
dual hairy black brane in asymptotically $AdS_5$ geometry,
and thus establish
that the conformal order in $\mathfrak{M}_{PW}^b$ CFT is unstable. 
\item We extend $\mathfrak{M}_{PW}^b$ holographic model when the boundary
CFT is compactified on $S^3$ of a radius $L$, equivalently, with the curvature
scale $K=\frac{1}{L^2}$. We show that the conformal order persists in the
limit $b\to \infty$ for a wide range of $\frac{K^{1/2}}{T}$. 
The ordered phase is subdominant both the canonical and the microcanonical
ensembles. It is always unstable: we identify a QNM mode of the
dual hairy black hole in asymptotically $AdS_5$ geometry with $\Im[\ww]>0$.
\item In the limit $b\to \infty$,
the ordered and the disordered phases become identical:
for the parameters introduced in \eqref{phd}, we find
$\kappa(b)-1\propto b^{-2}\to 0$ and
$\gamma(b)\propto b^{-1}\to 0$. 
The absence of the thermodynamically dominant 
conformal order in $\mathfrak{M}_{PW}^b$ holographic model in this limit
is consistent with the perturbative stability of the disordered phase,
as dictated by the general no-go theorem of section \ref{stable}. 
\item We study compactified conformal order in $\mathfrak{M}_{PW}^b$ model 
at finite $b$ for select values of $\frac{K^{1/2}}{T}$. We establish that
the conformal order exists only for $b>b_{crit}(K)$. Since $b_{crit}$ is found to be
always larger than 1, the holographic model  $\mathfrak{M}_{PW}^b$ with
the ordered phase  is always a deformation
of the $\caln=2^*$/Pilch-Warner top-down holography.
In other words, similar  to \cite{Buchel:2020xdk},
there is no conformal order in $\caln=2^*$ theory.
\item We establish the compactified conformal order in $\mathfrak{M}_{PW}^b$
model at finite $b$, when it exists, is thermodynamically
subdominant and unstable.
\item It is easy to 'symmetrize' the holographic model $\mathfrak{M}_{PW}^b$:
the resulting $\mathfrak{M}_{PW,sym}^b$ model has a $\zet_2$ parity symmetry 
$\calo_2\leftrightarrow -\calo_2$. In $\mathfrak{M}_{PW,sym}^b$ model,
the compactified  conformal order
is associated with the spontaneous breaking of this parity invariance.
All the features of this symmetric model are qualitatively identical to
those of  $\mathfrak{M}_{PW}^b$ model. 
\end{itemize}

We finish the introduction with the comment about the prospect for
constructing the thermodynamically dominant and
stable holographic conformal order. All the known
constructions of the holographic conformal order
\cite{Buchel:2009ge,Buchel:2020thm,Buchel:2020xdk}, as well as the models
discussed here, can be thought
to rely on a  perturbative  deformation of the disordered phase in the
holographic CFTs. The no-go theorem of section \ref{stable}
strongly suggests that one can never construct a thermodynamically dominant
phase in this fashion --- again we stress the difference here with the
QFT approaches of
\cite{Chai:2020zgq,Chaudhuri:2020xxb,Chai:2021djc,Chaudhuri:2021dsq}. 
So, if the stable holographic conformal order exists, it must be
discontinuous, in some sense, from the disordered phase.
The possible examples could be Klebanov-Strassler black branes/black holes
\cite{Buchel:2018bzp,Buchel:2021yay} 
in the limit of infinitely high temperature
$\frac{T}{\Lambda}\to \infty$,
where $\Lambda$ is the strong coupling scale of the
boundary cascading gauge theory. A big {\it if} is the existence
of these black branes/black holes in the high temperature limit
in the supergravity approximation. We expect to report on this
question in the near future.

\section{Summary}\label{sum}

The starting point of our analysis is an example
of the gauge theory/string theory holographic correspondence
between $\caln=2^*$ $SU(N)$ theory (in the planar limit
and at large 't Hooft coupling)  and type IIB supergravity
\cite{Pilch:2000ue,Buchel:2000cn,Buchel:2013id,Bobev:2013cja}.    
The corresponding bulk effective action resulting from the dimensional
reduction on the five-sphere of ten-dimensional supergravity takes form
\begin{equation}
S_5=\frac{1}{16\pi G_5}\int_{\calm_5} d^5\xi \sqrt{-g}\biggl[
R-12 (\del\alpha)^2-4(\del\chi)^2-V(\alpha,\chi)
\biggr]\,,
\eqlabel{pw}
\end{equation}
with the potential $V$ determined from the superpotential $W$
as
\begin{equation}
V=\frac 14\left(\frac 13\left(\frac{\del W}{\del \alpha}\right)^2
+\left(\frac{\del W}{\del \chi}\right)^2\right)-\frac 43 W^2\,,\qquad W=-e^{-2\alpha}-\frac 12
e^{4\alpha}\cosh(2\chi)\,.
\eqlabel{defv}
\end{equation}
The bulk scalars $\alpha$ and $\chi$ are (correspondingly)
the holographic dual to operators $\calo_2$ and $\calo_3$ of the
$\caln=4$ Yang-Mills \cite{Buchel:2000cn,Buchel:2007vy,Hoyos:2011uh}.
Finally, five-dimensional Newton's constant $G_5$
is\footnote{We set the asymptotic $AdS_5$ curvature radius to 2.}
\begin{equation}
G_5=\frac{4\pi}{N^2}\,.
\eqlabel{g5}
\end{equation}

The non-normalizable coefficients of the bulk scalars $\alpha$ and $\chi$
are related to the masses of the bosonic and fermionic components
of $\caln=2$ hypermultiplet. In this paper we are interested in
{\it conformal} theories, thus, these parameters are set to zero. 
To simplify the discussion, we can consistently set $\chi\equiv 0$,
leading to a conformal model  $\mathfrak{M}_{PW}$ --- $\caln=4$ $SU(N)$ SYM
with a single operator $\calo_2$ of dimension $\Delta=2$:
\begin{equation}
V_{\mathfrak{M}_{PW}}=-2 e^{2\alpha}-e^{-4\alpha}\,.
\eqlabel{vpw}
\end{equation}
Note that $\mathfrak{M}_{PW}$ does not have any global symmetry.
It does not have a conformal order either, as we already
alluded to in section \ref{intro}. 
Following \cite{Buchel:2020xdk}, we generate a class of deformed
conformal models $\mathfrak{M}_{PW}^b$, with 
\begin{equation}
\begin{split}
V_{\mathfrak{M}_{PW}^b}=&-3-12\alpha^2
+b\biggl(3+12\alpha^2-2 e^{2\alpha}-e^{-4\alpha}\biggr)
\\
=&-3 -12\alpha^2+8b\ \alpha^3+b\ \calo(\alpha^4)\,.
\end{split}
\eqlabel{vpwb}
\end{equation}
Note that
\begin{equation}
\lim_{b\to 1}\ V_{\mathfrak{M}_{PW}^b} =V_{\mathfrak{M}_{PW}}\,,
\eqlabel{pwbpert}
\end{equation}
\ie we recover in this limit the top-down holographic model $\mathfrak{M}_{PW}$.
We find that  there is a thermal ordered phase in $\mathfrak{M}_{PW}^b$,
provided $b>b_{crit}>1$. The precise value of $b_{crit}$ below which
the ordered phase ceases to exist depends on $K^{1/2}/T$, \ie the ratio
of the $S^3$ compactification scale and the temperature $T$.

In the rest of this section we explain the results only;
a curious reader can find technical details necessary to reproduce them in
section \ref{tech}.

\begin{figure}[t]
\begin{center}
\psfrag{k}[cc][][1][0]{$\frac{K^{1/2}}{T}$}
\psfrag{o}[bb][][1][0]{$\lim_{b\to \infty}\ b\cdot\langle\calo_2\rangle$}
\psfrag{w}[tt][][1][0]{$\lim_{b\to \infty}\ \Im[\ww]$}
\includegraphics[width=3in]{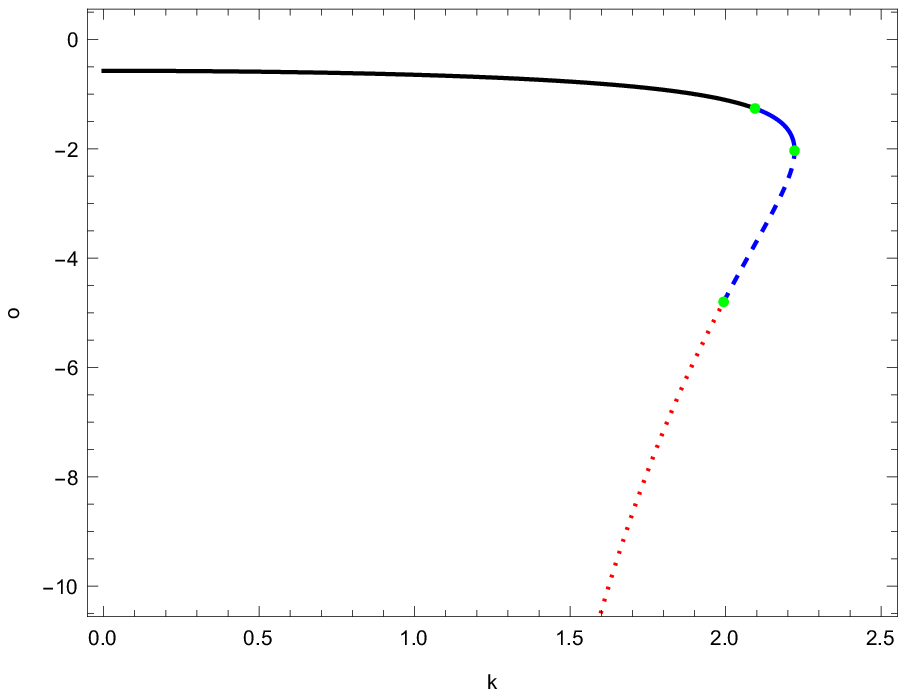}
\includegraphics[width=3in]{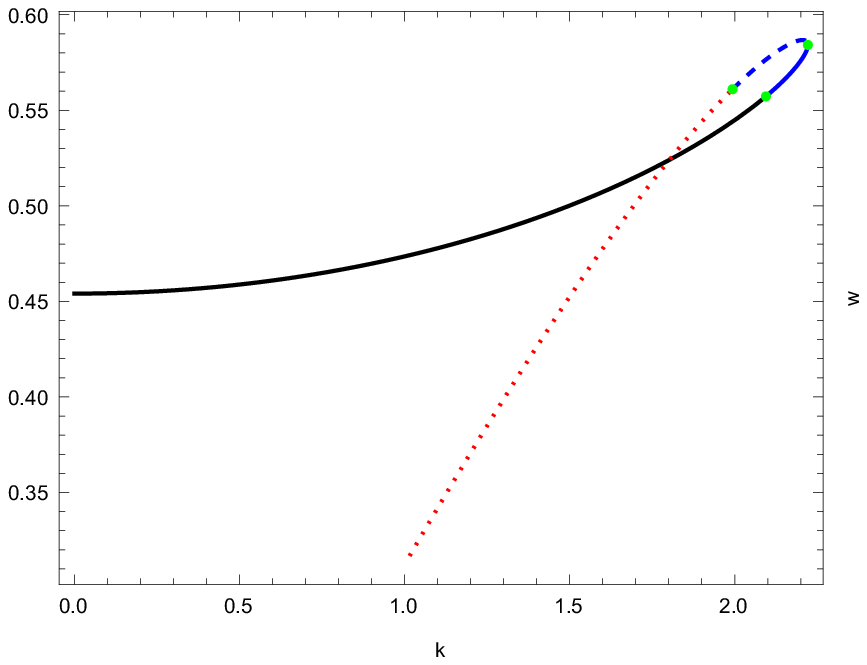}
\end{center}
  \caption{The left panel: the thermal expectation value of the
  order parameter $\calo_2$
  of the conformal model $\mathfrak{M}_{PW}^b$ in the limit
  $b\to +\infty$ as a function of $\frac{K^{1/2}}{T}$. The right panel:
  the unstable QNM of the hairy black hole representing
  the holographic dual to the ordered phase. 
} \label{O2}
\end{figure}

\begin{figure}[t]
\begin{center}
\psfrag{k}[cc][][1][0]{$\frac{K^{1/2}}{T}$}
\psfrag{e}[cc][][1][0]{$\frac{1}{\hat\cale}$}
\psfrag{f}[bb][][1][0]{$\lim_{b\to \infty}\ b^2\cdot\delta\hat{\calf}/\hat{T}^4 $}
\psfrag{s}[tt][][1][0]{$\lim_{b\to \infty}\ b^2\cdot\delta\hat{s}/\hat{\cale}^{3/4} $}
\includegraphics[width=3in]{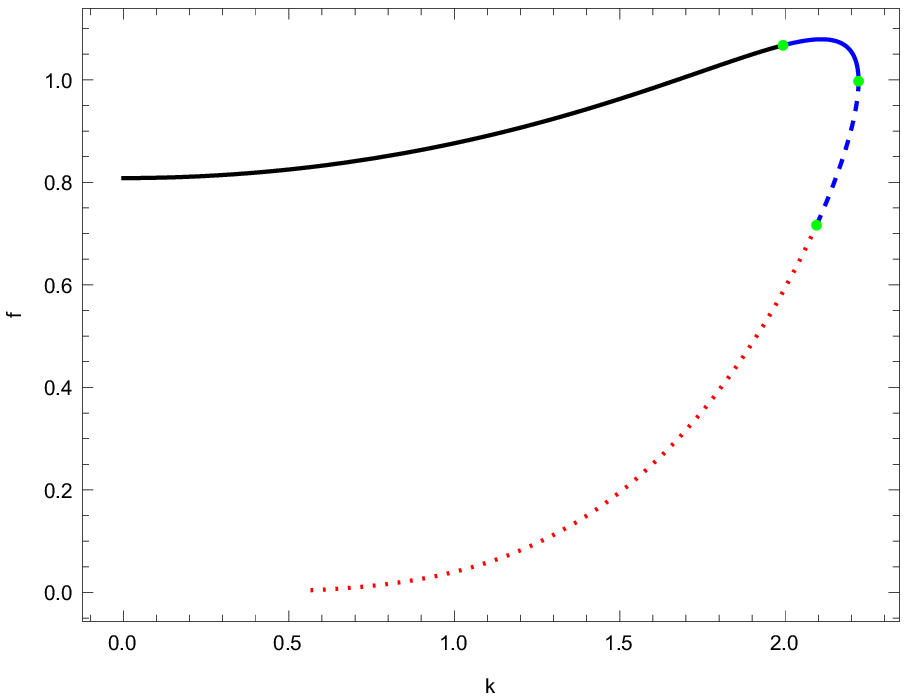}
\includegraphics[width=3in]{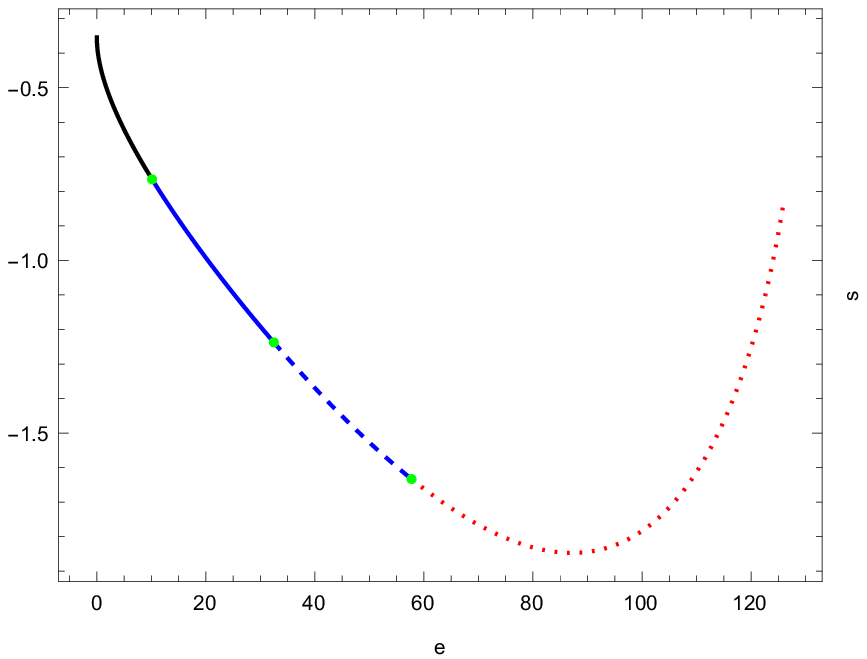}
\end{center}
  \caption{The ordered phase of the conformal model $\mathfrak{M}_{PW}^b$
  in the limit $b\to +\infty$ has a higher free energy density than that of
  the disordered phase at the corresponding temperature (the left panel).
  The ordered phase of the conformal model $\mathfrak{M}_{PW}^b$
  in the limit $b\to +\infty$ has a lower entropy density than that of
  the disordered phase at the corresponding energy density (the right panel).
  See \eqref{hats} and \eqref{defdeltas} for the definition of the
  reduced thermodynamic functions. 
} \label{ans}
\end{figure}

In fig.~\ref{O2} we present the results for the conformal order
in $\mathfrak{M}_{PW}^b$ model in the limit $b\to +\infty$ for different
values of $K^{1/2}/T$. Note that the thermal expectation value 
$\langle\calo_2\rangle\propto \frac 1b$ (the left panel),
implying that the backreaction
of the bulk scalar $\alpha$ on the geometry becomes vanishingly small
in the limit $b\to +\infty$.
Constructed conformal order is perturbatively unstable: in the right panel
we identify a QNM  of the corresponding hairy black hole with $\Im[\ww]>0$.
The disordered phase is represented holographically by $AdS_5$ black hole.
There are four distinct regimes of $K^{1/2}/T$ in the disordered phase,
which we also highlighted with solid/dashed/dotted curves in the
ordered phase:
\nxt the solid black curves correspond to
\begin{equation}
0<\frac{K^{1/2}}{T}<\frac {2\pi}{3}\,,
\eqlabel{reg1}
\end{equation}
with $AdS_5$ black holes temperature about the Hawking-Page transition
\cite{Hawking:1982dh,Witten:1998zw};
\nxt the solid blue curves correspond to
\begin{equation}
\frac {2\pi}{3}<\frac{K^{1/2}}{T}<\frac{\pi}{\sqrt{2}}\,,
\eqlabel{reg2}
\end{equation}
with $AdS_5$ black holes temperature below the Hawking-Page transition,
but having a positive specific heat;
\nxt the dashed blue curves correspond to
\begin{equation}
1.993(4)<\frac{K^{1/2}}{T}<\frac{\pi}{\sqrt{2}}\,,
\eqlabel{reg3}
\end{equation}
with $AdS_5$ black holes having a negative specific heat, but being
perturbatively stable with respect to localization on $S^5$
\cite{Hubeny:2002xn,Buchel:2015gxa};
\nxt the dotted red curves correspond to
\begin{equation}
\frac{K^{1/2}}{T}<1.993(4)\,,
\eqlabel{reg4}
\end{equation}
with $AdS_5$ black holes  being
perturbatively unstable with respect to localization on $S^5$.

In fig.~\ref{ans} we show that the ordered phase of the conformal model 
 $\mathfrak{M}_{PW}^b$  in the limit $b\to +\infty$ is subdominant
 both the in canonical ensemble (the left panel) and the microcanonical
 ensemble (the right panel).
 We define the reduced free energy density $\hat\calf$, the
 reduced energy density $\hat\cale$, the reduced entropy density
 $\hat s$, and the
 reduced temperature $\hat T$ as\footnote{The normalization
 is chosen so that $\lim_{\hat T\to \infty} \frac{\hat\calf}{\hat T^4}=-1$.} 
\begin{equation}
\begin{split}
&\hat\calf\equiv \frac{8}{\pi^2 N^2}\ \frac{\calf}{K^2}\,,\qquad
\hat\cale\equiv \frac{8}{\pi^2 N^2}\ \frac{\cale}{K^2}\,,\qquad
\hat s\equiv \frac{8}{\pi^2 N^2}\ \frac{s}{K^{3/2}}\,,\qquad
\hat T = \frac{T}{K^{1/2}}\,,
\end{split}
\eqlabel{hats}
\end{equation}
furthermore,
\begin{equation}
\delta\hat\calf\equiv \hat\calf_{ordered}-\hat\calf_{disordered}
\bigg|_{{\rm fixed} \ \hat T}\,,\qquad \delta\hat s\equiv \hat
s_{ordered}-\hat s_{disordered}
\bigg|_{{\rm fixed} \ \hat\cale}\,.
\eqlabel{defdeltas}
\end{equation}
The color/style coding  in the left panel is as in fig.~\ref{O2};
in the right panel the coding reflects the values of $\hat\cale$
of the disordered phase  
corresponding to \eqref{reg1}-\eqref{reg4}.

\begin{figure}[t]
\begin{center}
\psfrag{b}[cc][][1][0]{$b$}
\psfrag{w}[tt][][1][0]{$\Im[\ww]$}
\psfrag{f}[bb][][1][0]{$(\calf_{ord}-\calf_{disord})/|\calf_{disord}|$}
\includegraphics[width=3in]{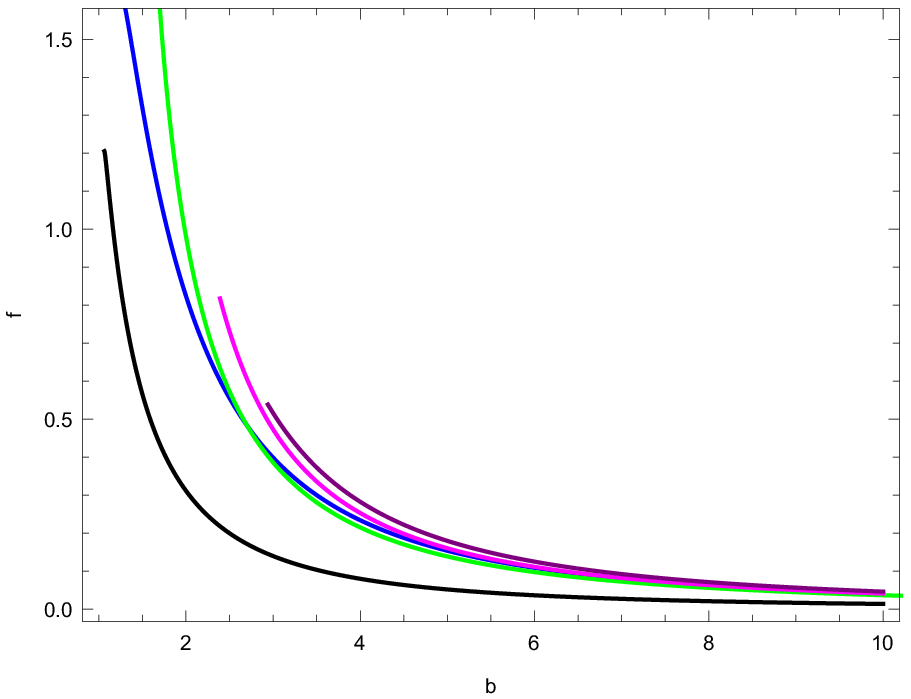}
\includegraphics[width=3in]{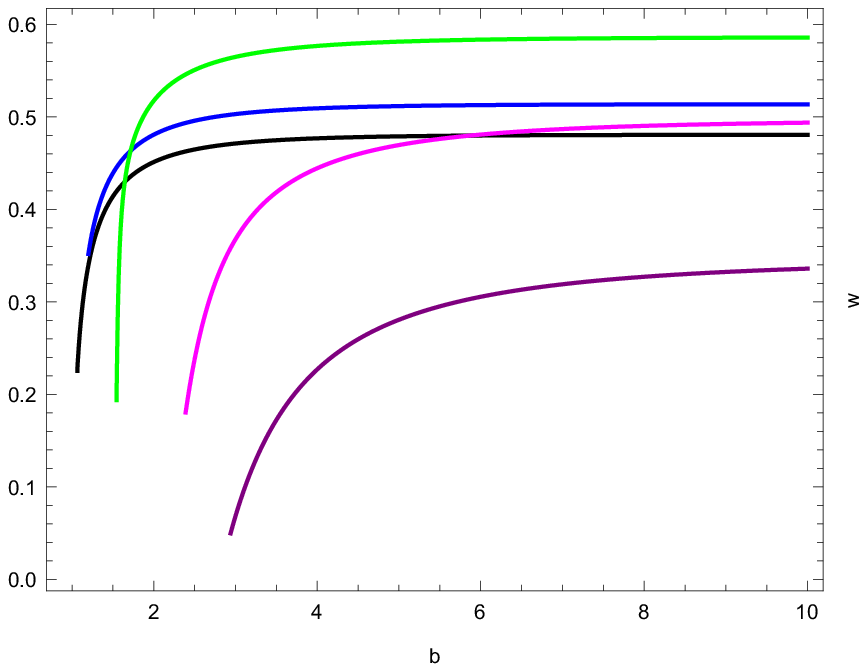}
\end{center}
  \caption{The ordered phase of the conformal model $\mathfrak{M}_{PW}^b$
at finite $b$ and select values of $K^{1/2}/T$, see \eqref{colork}.
The left panel demonstrates that the ordered phase has a higher free energy density
than that of the disordered phase (at the corresponding temperature).
The right panel identifies the QNM in hairy black holes, holographically
dual to the ordered phase, with $\Im[\ww]>0$, rendering this thermal ordered phase
perturbatively unstable.
} \label{sel}
\end{figure}

\begin{figure}[t]
\begin{center}
\psfrag{b}[cc][][1][0]{$b_{crit}$}
\psfrag{K}[tt][][1][0]{$\frac{K^{1/2}}{T}$}
\includegraphics[width=3in]{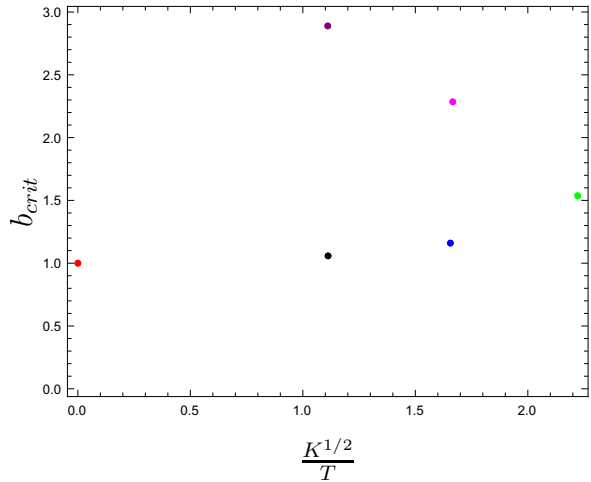}
\end{center}
  \caption{Thermal ordered phases in $\mathfrak{M}_{PW}^b$ model exist only for $b>b_{crit}$.
  $b_{crit}$ is estimated for the set of $K^{1/2}/T$ in \eqref{colork}. The red dot
  represents $b_{crit}(K=0)$.
} \label{bc}
\end{figure}

In fig.~\ref{sel} we present the results for the thermal ordered  
phase in $\mathfrak{M}_{PW}^b$ model for finite values of $b$ and select values of
$\frac{K^{1/2}}{T}$:
\begin{equation}
\frac{K^{1/2}}{T}\approx  \frac{\pi}{\sqrt{2}}\ \biggl\{{\color{black} \frac 12}\,,\,
{\color{blue} \frac 34}\,,\,
{\color{green} 1}\,,\,
{\color{magenta} \frac 34}\,,\,
{\color{purple} \frac 12}\biggr\} \,.
\eqlabel{colork}
\end{equation}
The black and the blue curves represent black holes with the positive specific heat;
the magenta and the purple curves represent black holes with the negative specific heat. 
The left panel indicates that the disordered phase has a lower free energy density,
and thus is the preferred one\footnote{The disordered phase is also the dominant one
in the microcanonical ensemble.}. The right panel demonstrates that the ordered phase
is perturbatively unstable: we identify the quasinormal mode in the helicity
zero sector of the dual hairy black hole with $\Im[\ww]>0$. 

Thermal ordered phases in $\mathfrak{M}_{PW}^b$ model exist only for $b>b_{crit}$.
In fig.~\ref{bc} we present the estimate for $b_{crit}$ for the set
of $K^{1/2}/T$ in \eqref{colork}. Additionally, the red dot indicates (see fig.~\ref{reddot})
\begin{equation}
b_{crit}\bigg|_{K=0}\approx 1.000(1)\,.
\eqlabel{bc0}
\end{equation}
The fact that $b_{crit}>1$ implies that there is no thermal ordered phase in
top-down holographic model $\mathfrak{M}_{PW}$.

\begin{figure}[t]
\begin{center}
\psfrag{k}[cc][][1][0]{$\frac{K^{1/2}}{T}$}
\psfrag{o}[bb][][1][0]{$\lim_{b\to \infty}\ \sqrt{b}\cdot|\langle\calo_2\rangle|$}
\psfrag{w}[tt][][1][0]{$\lim_{b\to \infty}\ \Im[\ww]$}
\includegraphics[width=3in]{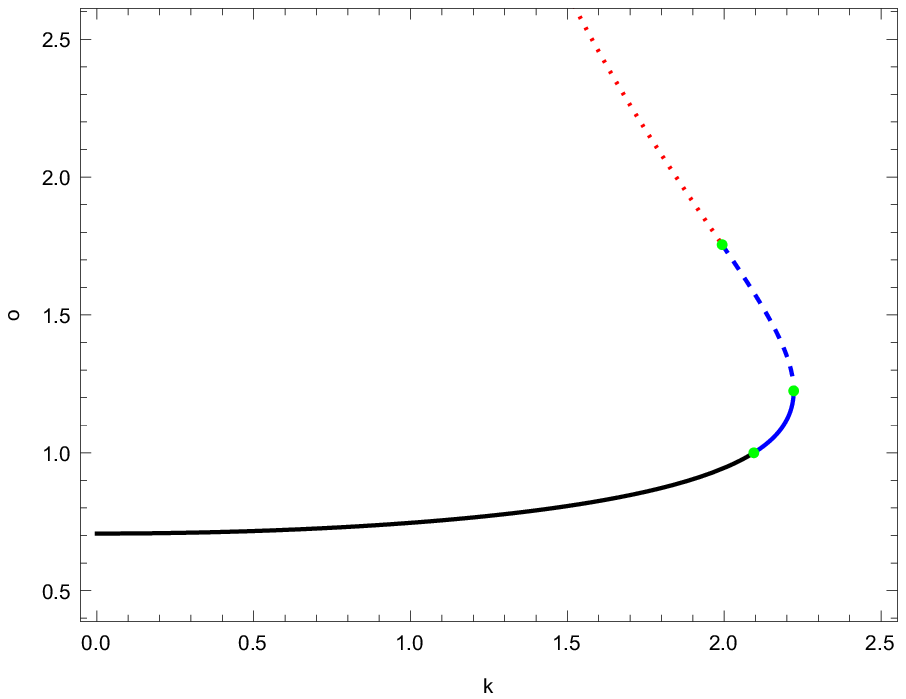}
\includegraphics[width=3in]{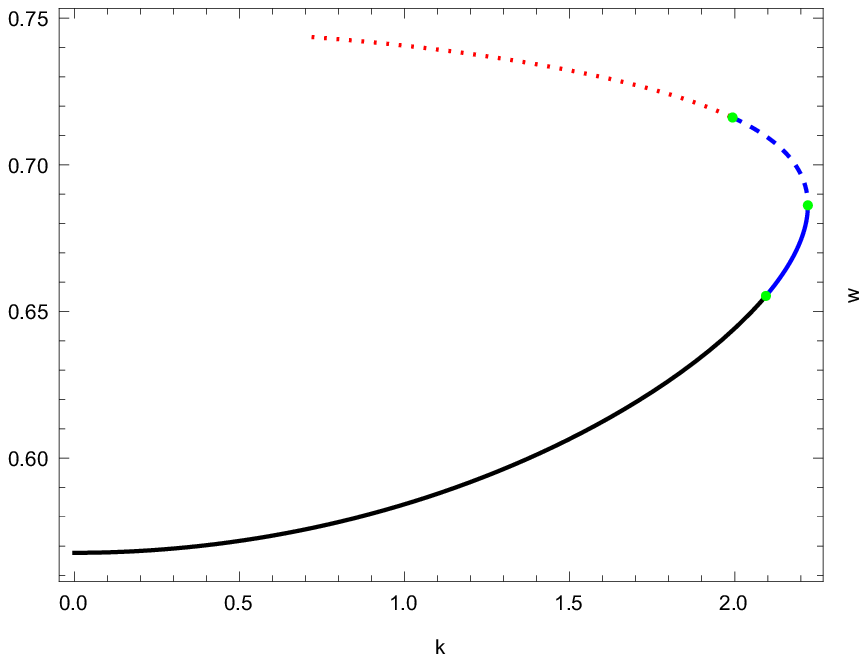}
\end{center}
  \caption{The left panel: the thermal expectation value of the
  order parameter $\calo_2$
  of the conformal model $\mathfrak{M}_{PW,sym}^b$ in the limit
  $b\to +\infty$ as a function of $\frac{K^{1/2}}{T}$. The right panel:
  the unstable QNM of the hairy black hole representing
  the holographic dual to the ordered phase. 
} \label{O2sym}
\end{figure}

So far we presented the evidence that the holographic conformal order survives the
compactification of the dual boundary theory on $S^3$, but remains unstable. 
In the example presented, the order was not associated with the spontaneous
breaking of any global symmetry. The reader might reasonable worry whether our conclusions
are specific to models without any global symmetry.
Rather, we claim that our results are generic: we extend 
analysis to $\mathfrak{M}_{PW,sym}^b$ model, where the bulk scalar potential
is $\zet_2$-symmetric,
\begin{equation}
\begin{split}
V_{\mathfrak{M}_{PW,sym}^b}=&\frac 12\ V_{\mathfrak{M}_{PW}^b}(\alpha)+\frac 12\ V_{\mathfrak{M}_{PW}^b}(-\alpha)
\\
=&-3 -12\alpha^2-12b\ \alpha^4-b\ \calo(\alpha^6)\,,
\end{split}
\eqlabel{vpwbs}
\end{equation}
and the conformal order is associated with the spontaneous breaking of this parity
invariance. In fig.~\ref{O2sym} we present the results for the conformal order
in $\mathfrak{M}_{PW,sym}^b$ model in the limit $b\to +\infty$ for different
values\footnote{The color/style coding is as in fig.~\ref{O2}.} of $K^{1/2}/T$.
The left panel shows the order parameter
$\langle\calo_2\rangle\propto \pm\frac {1}{\sqrt{b}}$, and the right panel
identifies an unstable QNM  of the corresponding hairy black hole.
The finite $b$ results of  $\mathfrak{M}_{PW,sym}^b$ model are qualitatively identical to
those of  $\mathfrak{M}_{PW}^b$ model.

\section{Stability of the holographic CFT disordered thermal states}
\label{stable}

In this section we prove that disordered phases of  four-dimensional
holographic CFTs on $\reals^{3}$ or $S^3$,
dual to bulk models of asymptotically $AdS_5$ Einstein gravity with
arbitrary scalars, are perturbatively
stable. These disordered phases are represented by AdS-Schwarzschild
black branes/black holes. It is well-know that AdS-Schwarzschild black branes/black
holes are stable with respect to the purely gravitational perturbations
\cite{Ishibashi:2003ap} --- we extend this statement to the stability
with respect to the bulk scalar fluctuations.

It is important to stress what potential instabilities are not covered
by the analysis below. A top-down conformal holographic model
is formulated in ten dimensional type IIB supergravity in asymptotically
$AdS_5\times \calv_5$, where $\calv_5$ is a compact manifold, $\calv_5=S^5$ in the
familiar example of
large-$N$ $\caln=4$ SYM. It is known that $AdS_5$-Schwarzschild black holes
can be unstable with respect to metric fluctuations carrying nonzero
momentum on $\calv_5$
\cite{Hubeny:2002xn,Buchel:2015gxa,Buchel:2015pla}. Once we reduce the
ten-dimensional holographic correspondence on $\calv_5$ we loose access
to such fluctuations. However, such instabilities are incorrect to interpret
as Schwarzschild black holes growing the scalar hair and violating the
no-hair theorem. Rather, these are the instabilities of the
initially smeared over $\calv_5$ black holes towards localization on the compact
transverse space.

Consider a holographic correspondence encoded in an  effective five-dimensional
gravitational action 
\begin{equation}
S_5=\frac{1}{16\pi G_5} \int_{\calm_5}d^5\xi\sqrt{-g}\biggl[
R-\frac 12\sum_{j=1}^p  (\del\phi_j)^2-V\left(\{\phi_j\}\right)
\biggr]\,.
\eqlabel{gen1}
\end{equation}
We will keep the scalar potential $V\left(\{\phi_j\}\right)$ arbitrary, in particular,
we will not assume any global symmetries in the model\footnote{
We explain shortly how the unitarity of the boundary $CFT_4$ constrains
this potential.}. To conform with the rest of the discussion in this paper,
we set the radius of the asymptotically $AdS_5$ geometry to 2, then,
the potential takes form 
\begin{equation}
V=\left(\{\phi_j\}\right)=-3-\frac 12 \sum_{i,j=1}^p m_{ij} \phi_i\phi_j
+\calo(\phi^3)\,.
\eqlabel{gen2}
\end{equation}
The disordered phase is the $AdS_5$-Schwarzschild black brane/black hole,
which we write in infalling Eddington-Finkelstein (EF) coordinates as
(compare with \eqref{background}) 
\begin{equation}
ds_5^2=-c_1^2\ d\tau^2-2 c_1 c_3\ d\tau dr+c_2^2\ d\Omega_{(3,K)}^2\,,
\eqlabel{metricEF}
\end{equation}
where $\tau$ is the EF time, and the minus sign in $d\tau dr$ term
is due to the fact that the AdS boundary is at $r\to 0$,  
with
\begin{equation}
\begin{split}
&c_1=\frac{f^{1/2}}{r h^{1/4}}\,,\qquad c_2=\frac{1}{r h^{1/4}}\,,\qquad
c_3=\frac{h^{1/4}}{r f^{1/2}}\,,\\
&f=\frac{(2 r+1) (16 K r^2+2 r^2+2 r+1)}{(r+1)^4}\,,\qquad h=\frac{16}{(1+r)^4}\,.
\end{split}
\eqlabel{fhads}
\end{equation}
Additionally, the background values of all the scalars are set to zero
\begin{equation}
\phi_j(r)\equiv 0\,.
\eqlabel{sads5}
\end{equation}
Following \cite{Buchel:2021ttt}, there are $p+3$ branches of the QNMs:
\nxt helicity $h=2$ and $h=1$ branches of the metric fluctuations;
\nxt $p+1$ branches of the helicity $h=0$ fluctuations.\\
$h=2$ and $h=1$ branches can not contain
instabilities \cite{Ishibashi:2003ap,Buchel:2021ttt},
thus we focus on the helicity $h=0$ fluctuations. Because the scalar fields
vanish in the background \eqref{sads5}, (the stable \cite{Ishibashi:2003ap})
helicity $h=0$
metric fluctuations will decouple from the fluctuations of the bulk scalars
(see eq.(A.22) of \cite{Buchel:2021ttt}). Moreover, for the same reason,
generically coupled $p$ branches of the scalar fluctuations
will depend only on the mass-matrix
coefficients $m_{ij}$ in the potential \eqref{gen2}, and not on the
details of nonlinearities $\calo(\phi^3)$  (see eq.(A.23) of \cite{Buchel:2021ttt}).
Thus, for all practical purposes we can truncate the generic scalar potential 
in \eqref{gen2} to $\calo(\phi^2)$. Using the orthogonal rotation 
in the field space, the equivalent to \eqref{gen1}
effective action (in regards to computing the
spectra of the QNMs) is given by
\begin{equation}
S_5=\frac{1}{16\pi G_5} \int_{\calm_5}d^5\xi\sqrt{-g}\biggl[
R+3-\frac 12\sum_{j=1}^p\biggl\{  (\del\chi_j)^2 + \mu_j^2\ \chi_j^2\biggr\}
\biggr]\,,
\eqlabel{gen3}
\end{equation}
with a new scalar $\chi_j$ of mass $\mu_j^2$ dual to an operator
$\calo_j$ of dimension $\Delta_j$ of the boundary CFT
\begin{equation}
\{\calo_j\,,\ \Delta_j\}\qquad \Longleftrightarrow\qquad \{\chi_j\,,\
4 \mu_j^2=\Delta_j(\Delta_j-4)\}\,,\qquad \Delta_j>1\,.
\eqlabel{gen4}
\end{equation}
In \eqref{gen4} we assumed the Breitenlohner-Freedman
bound \cite{Breitenlohner:1982jf} on the scalar masses, equivalently the
unitarity bound on the operators of the interactive boundary CFT
\cite{Klebanov:1999tb}.
An advantage of using \eqref{gen3} is that the branches of the scalar
fluctuations $\Phi_j$ now completely decouple:
\begin{equation}
0=\square \Phi_j-\mu_j^2\ \Phi_j=\square \Phi_j-\frac{\Delta_j(\Delta_j-4)}{4}\
\Phi_j\,,\qquad j=1,\cdots p\,.
\eqlabel{eomfl}
\end{equation}
QNM equations \eqref{eomfl} are solved with the regularity condition
at the horizon, and the boundary fall-off
\begin{equation}
\Phi_j\ \sim r^{\Delta_j}\,,\qquad r\to 0\,.
\eqlabel{gen5}
\end{equation}
Each of the $p$ QNM branches can be analyzed separately: from now on we drop
the subscript $ _j$ and consider $\Delta$ to be continuous with $\Delta>1$.

The argument below is standard in the literature \cite{Horowitz:1999jd} ---
we are simply being careful with the various boundary terms.
Introduce
\begin{equation}
\Phi=e^{-i w \tau}\ \frac{S(\Omega_{3,K})}{c_2^{3/2}}\ \psi(r)\,,\qquad {\rm where}\qquad 
\Delta_{\Omega_{3,K}}\ S + k^2\ S=0\,.
\eqlabel{depsi}
\end{equation}
From \eqref{eomfl} we find the second order equation for $\psi$:
\begin{equation}
0=\calx\equiv \left[\frac{c_1}{c_3}\ \psi'\right]'+2i\ w\ \psi' -c_1c_3\ V(r)\ \psi\,,
\eqlabel{eompsi}
\end{equation}
where using \eqref{fhads} we explicitly evaluate $V$ as
\begin{equation}
V=\frac{3 r^2 (4 r^2+2 r+1) K}{(r+1)^4}
+\frac{4 k^2 r^2}{(r+1)^2}+\frac{9r^4}{16(r+1)^4}+\frac{(2 \Delta-3) (2 \Delta-5)}{16}\,,
\eqlabel{defvads}
\end{equation}
with
\begin{equation}
k^2=\begin{cases}
&[0,+\infty)\,,\ {\rm if}\ K=0\,;\\
&K\ell (\ell +2)\,,\ \ell\in \zet_+\,,\ {\rm if}\ K>0 \,.
\end{cases}
\eqlabel{k2def}
\end{equation}
Eq.~\eqref{eompsi} needs to be solved subject to the boundary conditions
\begin{equation}
\begin{split}
&{\rm UV:}\qquad  \psi=r^{\Delta-3/2}\ \left(1+\calo(r)\right)\,,\qquad r\to 0\,, \\
&{\rm IR:}\qquad  \psi=\calo(1)\,,\qquad y\equiv \frac 1r\to 0 \,.
\end{split}
\eqlabel{bcpsi}
\end{equation}
In general, the solution of \eqref{eompsi} results in complex $\psi$ and $w$.

We present an analytic argument that $\Im[\ww]<0$ when $\Delta\ge \frac 52$;
for $\Delta\in (1,\frac 52)$ we need to resort to numerics. Establishing
$\Im[\ww]<0$ implies stability of the disordered phases of  holographic
CFTs with the effective action \eqref{gen1}. Note in particular that
$\mathfrak{M}_{PW}^b$ 
and $\mathfrak{M}_{PW,sym}^b$ models discussed in section \ref{sum}
are the special cases of \eqref{gen1}.  

\subsection{$\Delta\ge \frac 52$}\label{an}

Note that in this case the potential $V$ \eqref{eompsi}
is manifestly positive for $r\in (0,\infty)$.
Assuming $\{\psi,w\}$ is a solution to \eqref{eompsi}, we define
\begin{equation}
0=\cali(\epsilon,y_h)=-\int_{r=\epsilon}^{r=1/y_h} dr\ \calx\ \bar{\psi}\,.
\eqlabel{def}
\end{equation}
Ultimately, we will take the limit
\begin{equation}
0=\lim_{(\epsilon,y_h)\to 0}\ \cali(\epsilon,y_h)\equiv \cali\,.
\eqlabel{ilim}
\end{equation}
Using the explicit definition of $\calx$ \eqref{eompsi} and integrating by parts,
\begin{equation}
\cali(\epsilon,y_h)=\underbrace{
-\bar{\psi}\ \frac{c_1}{c_3}\ \psi'\bigg|_{r=\epsilon}^{r=1/y_h}}_{{\rm bt}}
+\underbrace{\int_{\epsilon}^{1/y_h} \biggl\{
\frac{c_1}{c_3}\ |\psi'|^2+c_1c_3\ V\ |\psi|^2
\biggr\} dr}_{\calj(\epsilon,y_h)}-2 i \underbrace{
w \int_{\epsilon}^{1/y_h}\psi' \bar{\psi} dr}_{\calh(\epsilon,y_h)}\,.
\eqlabel{def1}
\end{equation}
Note  that $\calj(\epsilon,y_h)$ is manifestly positive, and
the boundary term bt vanishes in the limit $(\epsilon,y_h)\to 0$
\begin{equation}
\begin{split}
&{\rm UV:}\qquad \bar{\psi}\ \frac{c_1}{c_3}\ \psi'\bigg|_{r=\epsilon}
\propto \left(\Delta-\frac 32\right) \epsilon^{2\Delta-4}\to 0\qquad {\rm since}\qquad
\Delta\ge \frac 52\,,
\\
&{\rm IR:}\qquad \bar{\psi}\ \frac{c_1}{c_3}\ \psi'\bigg|^{r=1/y_h}
\propto y_h\to 0\,.
\end{split}
\eqlabel{btlim}
\end{equation}
Thus
\begin{equation}
0=\cali=\calj-2 i \calh\qquad \Longrightarrow\qquad  \Re[\calh]=0\,,
\eqlabel{cont1}
\end{equation}
leading to
\begin{equation}
\begin{split}
0=&\int_{0}^{+\infty}\biggl\{w \psi' \bar{\psi}+ \bar{w} \bar{\psi}' \psi\biggr\} dr=
(w-\bar{w})\int_{0}^{+\infty} \psi' \bar{\psi}dr+\bar{w}\
|\psi|^2\bigg|_{r\to 0}^{r\to \infty}\\
=&(w-\bar{w})\int_{0}^{+\infty} \psi' \bar{\psi}dr+\bar{w} |\psi_h|^2\qquad
\Longrightarrow\qquad  \calh=\frac{|w|^2 }{-2 i \Im[w]}\ |\psi_h|^2\,,
\end{split}
\eqlabel{cont2}
\end{equation}
where we integrated the second term by parts and dropped
in the second line the $r\to 0$ boundary term since $|\psi|^2\propto
r^{2\Delta-3}\to 0$.  From \eqref{cont1} we find then
\begin{equation}
\int_{0}^{+\infty} \biggl\{
\frac{c_1}{c_3}\ |\psi'|^2+c_1c_3\ V\ |\psi|^2\biggr\} dr
=\ -\frac{|w|^2}{\Im[w]}\ |\psi_h|^2\qquad \Longrightarrow \qquad \Im[\ww]<0\,.
\eqlabel{cont3}
\end{equation}

\subsection{$\Delta\in(1,\frac 52)$}

The analytic proof of section \ref{an} fails here since:
\nxt the potential $V$
\eqref{defvads} can be negative for small $r$ when $\Delta\in (\frac 32,\frac 52)$;
\nxt the UV boundary term in \eqref{btlim} does not vanish for $\Delta\le 2$; 
\nxt the UV boundary term in \eqref{cont2} does not vanish for $\Delta\le \frac 32$.

Note that our models $\mathfrak{M}_{PW}^b$ and $\mathfrak{M}_{PW,sym}^b$
(and in particular the top-down example $\mathfrak{M}_{PW}$) have
disordered phase with $\Delta=2$. 
We will not attempt to find an analytic proof that $\Im[\ww]<0$
when $\Delta\in (1,\frac 52)$, and instead
present numerical evidence that this is indeed the case.

\begin{figure}[t]
\begin{center}
\psfrag{k}[cc][][1][0]{$\frac{K^{1/2}}{T}$}
\psfrag{r}[bb][][1][0]{$|\Re[\ww]|$}
\psfrag{i}[tt][][1][0]{$\Im[\ww]$}
\includegraphics[width=3in]{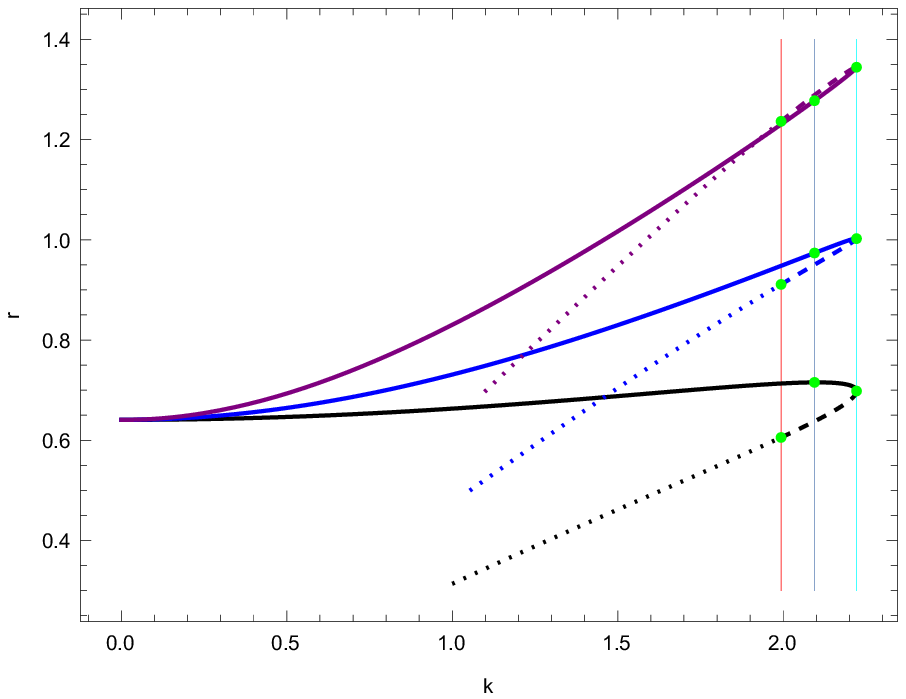}
\includegraphics[width=3in]{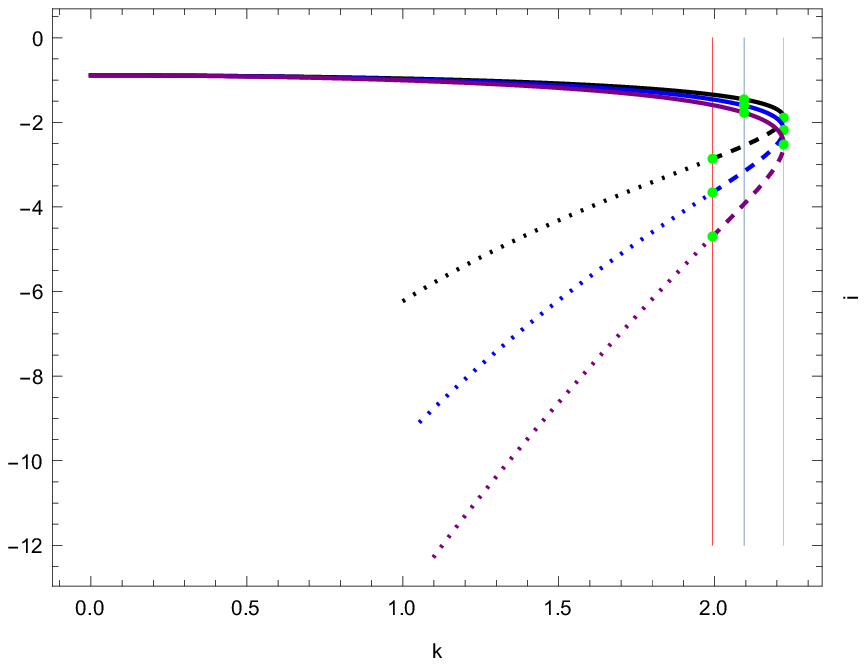}
\end{center}
  \caption{$\Re[\ww]$ (the left panel)
and $\Im[\ww]$ (the right panel) of the lowest $\Delta=2$ QNMs in the
disordered phase of $\mathfrak{M}_{PW}$ model
with $\ell=0$ (black curves), $\ell=1$ (blue curves)
and $\ell=2$ (purple curves).
} \label{delta2}
\end{figure}

\begin{figure}[t]
\begin{center}
\psfrag{d}[cc][][1][0]{$\Delta$}
\psfrag{r}[bb][][1][0]{$|\Re[\ww]|$}
\psfrag{i}[tt][][1][0]{$\Im[\ww]$}
\includegraphics[width=3in]{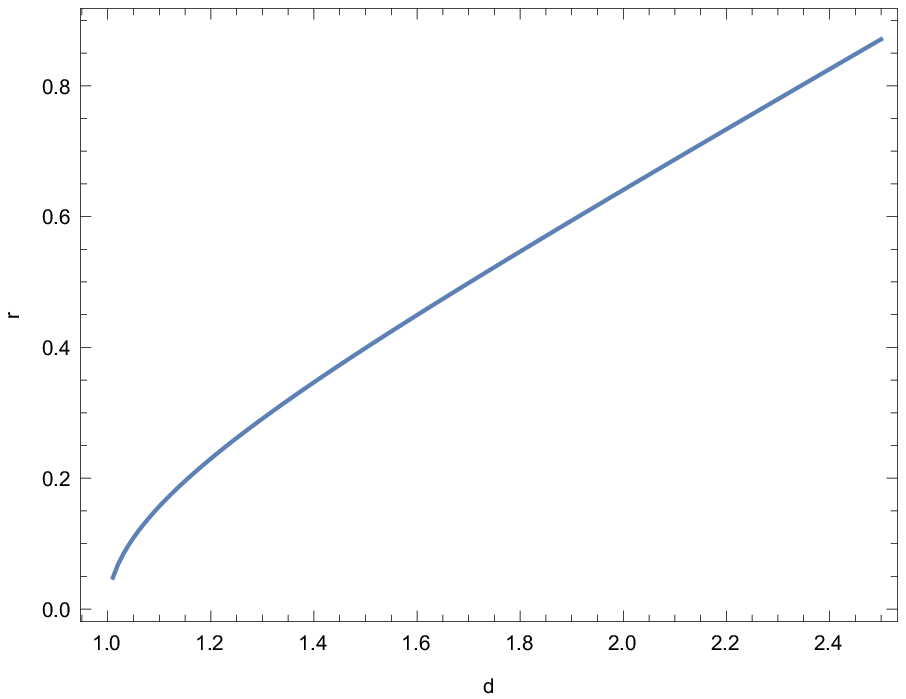}
\includegraphics[width=3in]{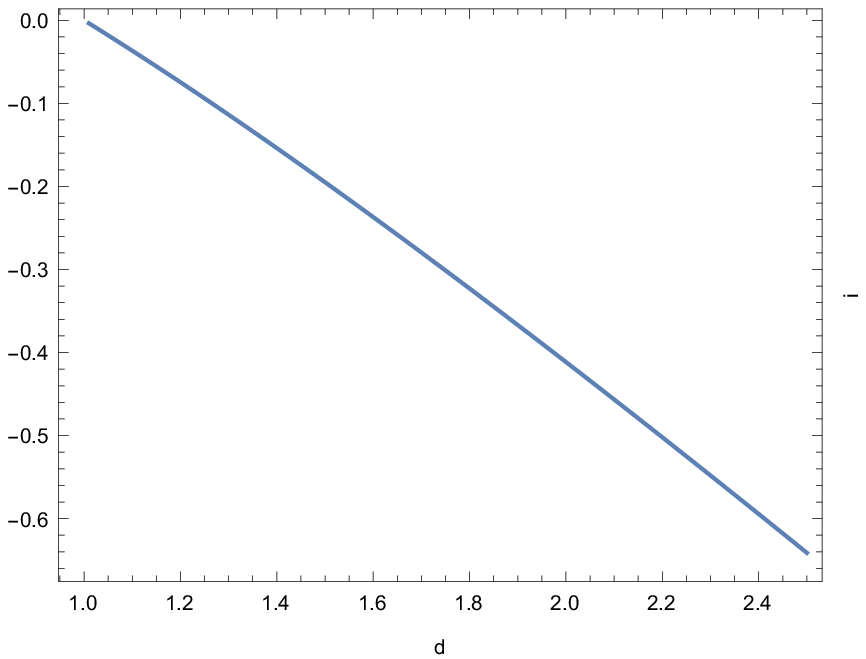}
\end{center}
  \caption{$\Re[\ww]$ (the left panel)
and $\Im[\ww]$ (the right panel) of the lowest $\ell=0$ QNMs in the
disordered phase with $\Delta\in (1,\frac{5}{2})$.
} \label{delta}
\end{figure}

In fig.~\ref{delta2} we present $\Re[\ww]$ (the left panel)
and $\Im[\ww]$ (the right panel) of the lowest $\Delta=2$ QNMs in the
disordered phase with $\ell=0$ (black curves), $\ell=1$ (blue curves)
and $\ell=2$ (purple curves). The vertical grey line
indicates the Hawking-Page transition of the $AdS_5$ Schwarzschild black hole;
the cyan line indicates the terminal temperature (the dashed/dotted black holes
have a negative specific heat), and the red line indicates the onset of the
localization instability of the $AdS_5\times S^5$ Schwarzschild black holes.

In fig.~\ref{delta} we present $\Re[\ww]$ (the left panel)
and $\Im[\ww]$ (the right panel) of the lowest $\ell=0$ QNMs in the
disordered phase with $\Delta\in (1,\frac{5}{2})$.

\section{Technical details}\label{tech}

In this section we highlight technical details useful to reproduce the
results reported in section \ref{sum}.

\subsection{$\mathfrak{M}_{PW}^b$ model in the limit $b\to +\infty$}

Conformal order in the limit $b\to +\infty$ becomes perturbative
\cite{Buchel:2020xdk}.  
We can solve eqs.~\eqref{eq1}-\eqref{eq3} as a series expansion in $\frac 1b$:
\begin{equation}
\begin{split}
&\alpha(r)=\sum_{n=1}^\infty \frac{1}{b^{2n-1}}\ \alpha_n(r)\,,\qquad
h=\frac{16}{(1+r)^4}\biggl(1+\sum_{n=1}^\infty
\frac{1}{b^{2n}}\ h_n(r)\biggr)\,,\\
&f=\frac{(2 r+1) (16 K r^2+2 r^2+2 r+1)}{(r+1)^4}\biggl(1+\sum_{n=1}^\infty
\frac{1}{b^{2n}}\ f_n(r)
\biggr)\,,
\end{split}
\eqlabel{pert1}
\end{equation}
where we explicitly factor the $AdS_5$-Schwarzschild black brane/black hole
solution in the strict $b\to +\infty$ limit, see \eqref{fhads}. 
To the leading $n=1$ order we find:
\begin{equation}
\begin{split}
0=&\alpha_1''+\frac{32 K r^4+4 r^4-16 K r^2-10 r^2-10 r-3}{r (2 r+1) ((16 K+1) r^2+ (r+1)^2) (r+1)}
\alpha_1'\\
&-\frac{4 \alpha_1 (r+1)^2 (\alpha_1-1)}{((16 K+1) r^2+ (r+1)^2) (2 r+1) r^2}\,,
\end{split}
\eqlabel{pert2}
\end{equation}
\begin{equation}
\begin{split}
0=&h_2''+\frac{2}{r+1} h_2'-16 (\alpha_1')^2\,,
\end{split}
\eqlabel{pert3}
\end{equation}
\begin{equation}
\begin{split}
0=&f_2'+4 (r+1) r (\alpha_1')^2
-\frac{8 K r^4+48 K r^3+r^4+24 K r^2+8 r^3+12 r^2+8 r+2}{(2 r+1) (16 K r^2+2 r^2+2 r+1)}
h_2'
\\&-\frac{4 (r+1) (8 K r^2+r^2+2 r+1)}{(2 r+1) (16 K r^2+2 r^2+2 r+1) r} f_2
+\frac{2 (r+1) (16 K r^2+r^2+2 r+1)}{(2 r+1) (16 K r^2+2 r^2+2 r+1) r} h_2
\\&-\frac{16(r+1)^3 (2 \alpha_1-3) \alpha_1^2}{3r (2 r+1) (16 K r^2+2 r^2+2 r+1)}\,.
\end{split}
\eqlabel{pert4}
\end{equation}
Eqs.~\eqref{pert2}-\eqref{pert4} are solved with the following asymptotics:
\nxt in the UV, \ie as $r\to 0$
\begin{equation}
\alpha_1=\alpha_{1,2} r^2 +\calo(r^3)\,,\qquad
h_2=\frac{16}{3} \alpha_{1,2}^2 r^4+\calo(r^5)\,,\qquad f_2=f_{2,4} r^4+\calo(r^5)\,,
\eqlabel{uvpert}
\end{equation}
\nxt in the IR, \ie as $y\equiv \frac 1r\to 0$
\begin{equation}
\begin{split}
&\alpha_1=\alpha_{1,0}^h +\calo(y)\,,\qquad
h_2=h_{2,0}^h+h_{2,1}^hy+\calo(y^2)\,,\\
&f_2=\frac{16 (\alpha_{1,0}^h)^2 (3-2 \alpha_{1,0}^h)+6 h_{2,0}^h (16 K+1)+3 h_{2,1}^h
(8 K+1)}{12(8 K+1)}+\calo(y)\,.
\end{split}
\eqlabel{irpert}
\end{equation}

From \eqref{thermores} we find
\begin{equation}
\begin{split}
&\hat\cale=\frac{3(8 K+1)^2}{256\pi^4 K^2}
-\frac{3f_{2,4}}{256\pi^4 K^2}\ \frac{1}{b^2}+\calo(b^{-4})\,,\ \ 
\hat s=\frac{1}{16 \pi^3 K^{3/2}}-\frac{3 h_{2,0}^h}{64\pi^3 K^{3/2}}\
\frac{1}{b^2}+\calo(b^{-4})\,,\\
&\hat T=\frac{8 K+1}{4\pi K^{1/2}}
+ \frac{16 (\alpha_{1,0}^h)^2 (3-2 \alpha_{1,0}^h)+48 K h_{2,0}^h
+3 h_{2,1}^h (8 K+1)}{48\pi K^{1/2}}\
\frac{1}{b^2}+\calo(b^{-4})\,.
\end{split}
\eqlabel{thermopert}
\end{equation}
The first law of thermodynamics at order $\calo(b^{-2})$ leads to the
constraint
\begin{equation}
\begin{split}
&0=\delta_{pert}\equiv 1-\frac{2 K (8 K+1)}{h_{2,0}^h (3+8 K)}\ (h_{2,0}^h)'
+\frac{2 K}{h_{2,0}^h (3+8 K)}\ (f_{2,4})'-\frac{8 K+1}{h_{2,0}^h (3+8 K)}\
h_{2,1}^h
\\&-\frac{4}{h_{2,0}^h (3+8 K)}\ f_{2,4}+\frac{16 (\alpha_{1,0}^h)^2 (2 \alpha_{1,0}^h-3)}{3h_{2,0}^h (3+8 K)}\,,
\end{split}
\eqlabel{pertconst}
\end{equation}
where $ '$ stands for $\frac{d}{dK}$. We verify the first law of thermodynamics of the
conformal order at $\calo(b^{-2})$ in fig.~\ref{error}.

We now discuss the computation of the QNMs in the perturbative conformal order. 
In the limit $b\to +\infty$, the differential operator
$\cald_2$ in \eqref{defcald} is
\begin{equation}
\cald_2=\cald_2\bigg|_{AdS_5-{\rm Schwarzschild}}+\calo(b^{-2})\,,
\eqlabel{d2limit}
\end{equation}
while $W$ in \eqref{w00pw} is
\begin{equation}
W=-1+{\mathcolorbox{red}{2\alpha_1}}+\calo(b^{-2})\,.
\eqlabel{wlimit}
\end{equation}
Note that the highlighted term in $W$ implies that the QNM spectra
of the ordered and the disordered phases are different even in the strict
$b\to +\infty$ limit.

\subsection{$\mathfrak{M}_{PW}^b$ model at finite $b$}

Using the equations of motions and the asymptotics for the
background and the QNMs in appendices \ref{apa} and \ref{apb},
the numerical solution is routine. We use the computation
techniques developed in \cite{Aharony:2007vg}.
To validate  numerical solutions we verify the first law \eqref{firstlaw}.
For example, for $K=0$, the first law can be stated as
\begin{equation}
0=\delta_{K=0}\equiv \frac{4\hat\cale-3  \hat T \hat s}{4\hat\cale}-1\,.
\eqlabel{firstlaw2}
\end{equation}
We present the results for $\delta_{K=0}$ in fig.~\ref{error}.

\begin{figure}[t]
\begin{center}
\psfrag{b}[cc][][1][0]{$b$}
\psfrag{h}[bb][][1][0]{$(h_0^h)^{-1}$}
\psfrag{g}[tt][][1][0]{$(h_0^h)^{-1}$}
\includegraphics[width=3in]{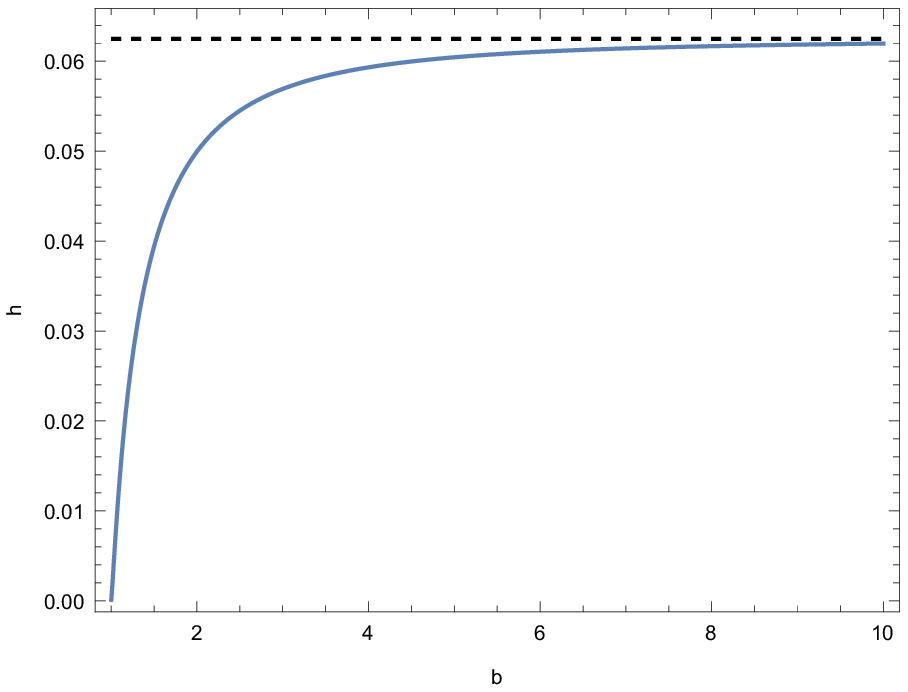}
\includegraphics[width=3in]{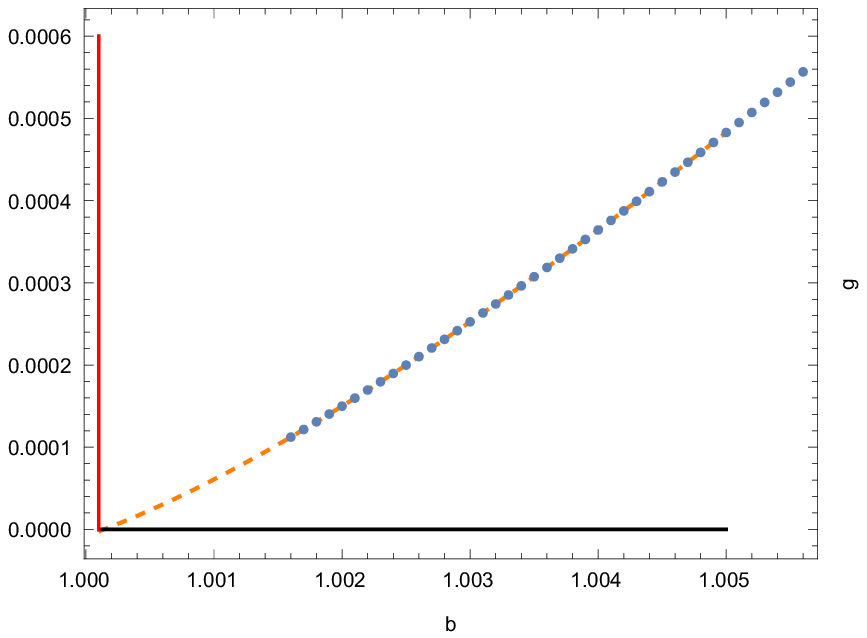}
\end{center}
  \caption{Computation of $b_{crit}$ in $\mathfrak{M}_{PW}^b$ model
  at $K=0$. The IR parameter $h_0^h$ (see \eqref{ir}) determining the bulk
  solution diverges as $b\to b_{crit}$. We estimate the location of this
  divergence interpolating the small-$b$ data (the dashed orange curve in
  the right panel). 
} \label{reddot}
\end{figure}

We conclude this section explaining how to obtain the results for $b_{crit}$
reported in fig.~\ref{bc}. We focus on the red dot in the figure, representing
the uncompactified ($K=0$) conformal order in $\mathfrak{M}_{PW}^b$ model.
The background geometry representing the conformal order is determined
by the set of parameters in the asymptotic expansions --- see
\eqref{uv1}-\eqref{ir}. Consider the parameter $h_0^h$. In the limit
$b\to +\infty$ we can use the $AdS_5$-Schwarzschild solution \eqref{fhads}
to conclude (remember the redefinition \eqref{defhh})
\begin{equation}
\lim_{b\to +\infty} \left(h_0^h\right)^{-1} = \frac {1}{16}\,.
\eqlabel{assh}
\end{equation}
This asymptote is represented by a dashed black line in the left panel of
fig.~\ref{reddot}. As $b$ decreases, $h_0^h$ increases, ultimately diverging
as $b\to b_{crit}$. Of course we can not reliably construct singular solutions ---
in the right panel of fig.~\ref{reddot} the points indicate values of
$\left(h_0^h\right)^{-1}$ for small $b$. We use Wolfram Mathematica
interpolation (orange dashed curve) to estimate the location of the
divergence of $h_0^h$
(highlighted with the vertical red line).

\subsection{$\mathfrak{M}_{PW,sym}^b$ model in the limit $b\to +\infty$}

The perturbative expansion in $\mathfrak{M}_{PW,sym}^b$ model takes form
\begin{equation}
\begin{split}
&\alpha(r)=\sum_{n=1}^\infty \frac{1}{b^{n-1/2}}\ \alpha_n(r)\,,\qquad
h=\frac{16}{(1+r)^4}\biggl(1+\sum_{n=1}^\infty
\frac{1}{b^{n}}\ h_n(r)\biggr)\,,\\
&f=\frac{(2 r+1) (16 K r^2+2 r^2+2 r+1)}{(r+1)^4}\biggl(1+\sum_{n=1}^\infty
\frac{1}{b^{n}}\ f_n(r)
\biggr)\,.
\end{split}
\eqlabel{pert1s}
\end{equation}
To reproduce the results reported in fig.~\ref{O2sym} we need the equation
for $\alpha_1$, and the leading order expression for  $W$
(computing the spectrum of the QNMs as in \eqref{qnm2}):
\begin{equation}
\begin{split}
0=&\alpha_1''+\frac{16 K r^2(2r^2-1)+4 r^4-10 r^2-10 r-3}{r (2 r+1)
(16 K r^2+2 r^2+2 r+1) (r+1)}\  \alpha_1'\\
&+\frac{4 \alpha_1 (r+1)^2 (2\alpha_1^2+1)}{
(16 K r^2+2 r^2+2 r+1) (2 r+1) r^2}\,,\\
&\alpha_1=\alpha_{1,2} r^2+\calo(r^3)\,,\qquad \alpha_1=\alpha_{1,0}^h+\calo(y)\,,
\end{split}
\eqlabel{a1sym}
\end{equation}
and 
\begin{equation}
W=-1-6\alpha_1^2+\calo(b^{-1})\,.
\eqlabel{Wsym}
\end{equation}

\section*{Acknowledgments}
This research is supported in part by Perimeter Institute for
Theoretical Physics.  Research at Perimeter Institute is supported in
part by the Government of Canada through the Department of Innovation,
Science and Economic Development Canada and by the Province of Ontario
through the Ministry of Colleges and Universities. This work was
further supported by NSERC through the Discovery Grants program.

\appendix

\section{Hairy black holes and their thermodynamics in $\mathfrak{M}_{PW}^b$ model}
\label{apa}

We parameterize black hole solutions in $\mathfrak{M}_{PW}^b$ model as
\begin{equation}
ds_5^2=-c_1^2\ dt^2+c_2^2\ d\Omega_{(3,K)}^2+c_3^2\ dr^2\,,\qquad \alpha=\alpha(r) \,,
\eqlabel{background}
\end{equation}
where the radial coordinate $r$ ranges as
\begin{equation}
r\in (0,+\infty)\,,
\eqlabel{rrange}
\end{equation}
where $c_i=c_i(r)$, 
\begin{equation}
c_1=\frac{f^{1/2}}{r h^{1/4}}\,,\qquad c_2=\frac{1}{r h^{1/4}}\,,\qquad
c_3=\frac{h^{1/4}}{r f^{1/2}}\,,
\eqlabel{warps}
\end{equation}
and the round 3-sphere metric $d\Omega_{(3,K)}^2$ of radius $K^{-1/2}$ is
\begin{equation}
d\Omega_{3,K}^2=\frac{dx^2}{(1-K x^2)}+(1-K x^2)\ \biggl[\frac{dy^2}{(1-K y^2)}+(1-K y^2)\ dz^2\biggr]\,.
\eqlabel{dx3k}
\end{equation}
From the gravitational effective Lagrangian of $\mathfrak{M}_{PW}^b$ model
(see \eqref{vpwb} for the scalar potential)
\begin{equation}
\call_{\mathfrak{M}_{PW}^b}=R-12(\del\alpha)^2-V_{\mathfrak{M}_{PW}^b}\,,
\end{equation}
we obtain the following second order equations:
\begin{equation}
\begin{split}
&0=f''-\frac{3f'}{r} -\frac{5h'f'}{4h}+4 h K\,,
\end{split}
\eqlabel{eq1}
\end{equation}
\begin{equation}
\begin{split}
&0=h''-\frac{5(h')^2}{4h}-16 h (\alpha')^2\,,
\end{split}
\eqlabel{eq2}
\end{equation}
\begin{equation}
\begin{split}
&0=\alpha''+a' \biggl(
\frac{f'}{f}-\frac3r-\frac{5h'}{4h}
\biggr)+\frac{h^{1/2}b}{6r^2 f} \left(e^{2 \alpha}-e^{-4 \alpha}\right)+(1-b)
\frac{h^{1/2}\alpha}{r^2 f}\,,
\end{split}
\eqlabel{eq3}
\end{equation}
and the first order constraint
\begin{equation}
\begin{split}
0=&(\alpha')^2+\frac{h K}{2f}-\frac{(h')^2}{16h^2}
+\frac{h' f'}{16f h}-\frac{h'}{2r h}+\frac{f'}{4r f}-\frac{1}{r^2}
+\frac{h^{1/2} b}{6r^2 f} \left(e^{2 \alpha}+\frac12 e^{-4 \alpha}\right)\\
&+(1-b)\ \frac{h^{1/2}\alpha^2}{r^2 f}\,.
\end{split}
\eqlabel{eqc}
\end{equation}
Eqs.~\eqref{eq1}-\eqref{eqc} are solved with the following asymptotics:
\nxt in the UV, \ie as $r\to 0$
\begin{equation}
\begin{split}
&f=1+16 K r^2-32 K r^3+f_4 r^4+\calo(r^5)\,,
\end{split}
\eqlabel{uv1}
\end{equation}
\begin{equation}
\begin{split}
&h=16-64 r+160 r^2-320 r^3+\left(560+\frac{256}{3} a_2^2\right) r^4+\calo(r^5)\,,
\end{split}
\eqlabel{uv2}
\end{equation}
\begin{equation}
\begin{split}
&\alpha=a_2 r^2-2 a_2 r^3+(a_2^2 b+3 a_2) r^4+\calo(r^5)\,,
\end{split}
\eqlabel{uv3}
\end{equation}
\nxt in the IR, \ie as $y\equiv \frac 1r\to 0$
\begin{equation}
\begin{split}
&f=f_1^h y+\calo(y^2)\,,\qquad \hat{h}=h_0^h+\calo(y)\,,\qquad 
\alpha=a_0^h+\calo(y)\,,
\end{split}
\eqlabel{ir}
\end{equation}
where we defined
\begin{equation}
\hat{h}\equiv y^{-4}\ h \,.
\eqlabel{defhh}
\end{equation}

Following the holographic renormalization of the related $\caln=2^*$ model 
\cite{Buchel:2015lla,Buchel:2004hw,Buchel:2012gw} we find:
\begin{equation}
\begin{split}
&\hat{\cale}=\frac{1}{8\pi^4}\biggl(6+\frac{9}{2K}
-\frac{3f_4}{32K^2}\biggr)\,,\qquad
\hat{s}=\frac{1}{2\pi^3(h_0^h)^{3/4} K^{3/2}}\,,\qquad
\hat{T}=\frac{f_1^h}{4\pi (h_0^h)^{1/2} K^{1/2}}\,.
\end{split}
\eqlabel{thermores}
\end{equation}
The basic thermodynamic relation,
\begin{equation}
\hat\calf=\hat\cale-\hat s \hat T\,,
\eqlabel{fes}
\end{equation}
is automatically enforced by the holographic renormalization
\cite{Buchel:2004hw}, while the first law of thermodynamics,
\begin{equation}
d\hat\cale=\hat T\ d\hat s\ \bigg|_{b={\rm const}}\,,
\eqlabel{firstlaw}
\end{equation}
must be verified numerically. We always check \eqref{firstlaw}
in numerical constructions of  $\mathfrak{M}_{PW}^b$ model black hole
geometries. See appendix \ref{sample} for a sample of such tests.

\section{$h=0$ QNMs of the hairy black holes in $\mathfrak{M}_{PW}^b$ model}
\label{apb}

We follow the framework of \cite{Buchel:2021ttt} for the computation of the QNMs.
We will be interested in the helicity $h=0$ quasinormal modes with $\ell=0$.
Note that at $\ell=0$ the metric and the bulk scalar fluctuations decouple.
We use  $F(t,r)$ to denote gauge invariant fluctuations associated with
the bulk scalar $\alpha$ of the conformal model $\mathfrak{M}_{PW}^b$.
Using the background parameterization
\eqref{warps}, we obtain from \cite{Buchel:2021ttt}:
\begin{equation}
\begin{split}
0=&\cald_2 F\  -W\ F\,,\\
\end{split}
\eqlabel{qnm2}
\end{equation}
where the second-order differential operator $\cald_2$ (coming
from $\square$ on the background geometry \eqref{background}) is
(note that $k^2=K \ell (\ell+2)=0$)
\begin{equation}
\begin{split}
\cald_2 F(t,r)&\equiv  -\frac{h^{1/2} r^2}{f}\ \del^2_{tt} F+\frac{r^2 f}{h^{1/2}}\
\del^2_{rr}F+\biggl(
\frac{r^2f'}{h^{1/2}}-\frac{5r^2 f h'}{4h^{3/2}}-\frac{3 r f}{h^{1/2}}
\biggr)\ \del_r F-{h^{1/2} r^2 k^2 }\ F\\
&=-\frac{h^{1/2} r^2}{f}\ \del^2_{tt} F+\frac{r^2 f}{h^{1/2}}\
\del^2_{rr}F+\biggl(
\frac{r^2f'}{h^{1/2}}-\frac{5r^2 f h'}{4h^{3/2}}-\frac{3 r f}{h^{1/2}}
\biggr)\ \del_r F\,,
\end{split}
\eqlabel{defcald}
\end{equation}
and
\begin{equation}
\begin{split}
&W=\frac{1}{3h G^2} \biggl(
1536 h^{5/2} f r^4 (\alpha')^4
+96 h^{1/2} r^2 (h' r+4 h)
(h f' r-4 h f-h' f r) (\alpha')^2\\
&+32 \alpha' (h' r+4 h) \left(6 \alpha h^2 r (1-b)+h^2 r
b (e^{2 \alpha}-e^{-4 \alpha})\right)
\biggr)
-\frac b3 \left( e^{2 \alpha}+2 e^{-4 \alpha}\right)+b-1\,,
\end{split}
\eqlabel{w00pw}
\end{equation}
with 
\begin{equation}
G\equiv h' r+4 h\,.
\eqlabel{defG}
\end{equation}

Generically, $F$, as well as $\ww$, are complex.
We need to impose the normalizable boundary conditions as $r\to 0$,
and the incoming wave boundary conditions at the black brane/black hole
horizon, \ie as $y\equiv \frac 1r\to 0$. We can explicitly factor
the boundary conditions, and the harmonic time dependence,
redefining $F$ as
\begin{equation}
\begin{split}
&F(t,r)=(1+r)^{i \ww/2}\ \frac{r^2}{1+r^2}\ e^{-i 2\pi T\ww t}\ f(r)\,,
\end{split}
\eqlabel{redeffs}
\end{equation}
which renders $f(r)$ 
regular and $\calo(1)$ both at the boundary and at the horizon.

Turns out that the unstable QNMs have $\Im[f]=0$ and $\Re[\ww]=0$.
Further introducing
\begin{equation}
f=f_{\Re}\,,\qquad \ww=i\ \ww_{\Im}\,,
\eqlabel{splitrei}
\end{equation}
we obtain from \eqref{qnm2} a single second order linear ODEs for
the real function $f_{\Re}$.
Because of the linearity, there is an arbitrary  overall
normalization of a solution; we fix this normalization imposing
\begin{equation}
\begin{split}
\lim_{r\to 0} f_{\Re}=1\,.
\end{split}
\eqlabel{normalization}
\end{equation}
The QNM equation for $f_{\Re}$ is solved with the following asymptotics:
\nxt in the UV, \ie as $r\to 0$
\begin{equation}
\begin{split}
&f_{\Re}=1+\left(-2+\frac12 \ww_\Im\right)\ r+\calo(r^2)\,,
\end{split}
\eqlabel{qnmuv}
\end{equation}
\nxt in the IR, \ie as $y\equiv \frac 1r\to 0$
\begin{equation}
\begin{split}
&f_{\Re}=f_{\Re,0}^h+\calo(y)\,.
\end{split}
\eqlabel{qnmir}
\end{equation}
Note that, for a fixed background,
the solution is characterized in total by 2 parameters
\begin{equation}
\left\{\ \ww_\Im\,,\ f_{\Re,0}^h\ \right\}\,,
\eqlabel{parqnm}
\end{equation}
precisely as needed to specify a solution of a single second order ODE.

\section{Numerical tests}\label{sample}

\begin{figure}[t]
\begin{center}
\psfrag{k}[cc][][1][0]{$K$}
\psfrag{b}[cc][][1][0]{$b$}
\psfrag{e}[bb][][1][0]{$\delta_{pert}$}
\psfrag{d}[tt][][1][0]{$\delta_{K=0}$}
\includegraphics[width=3in]{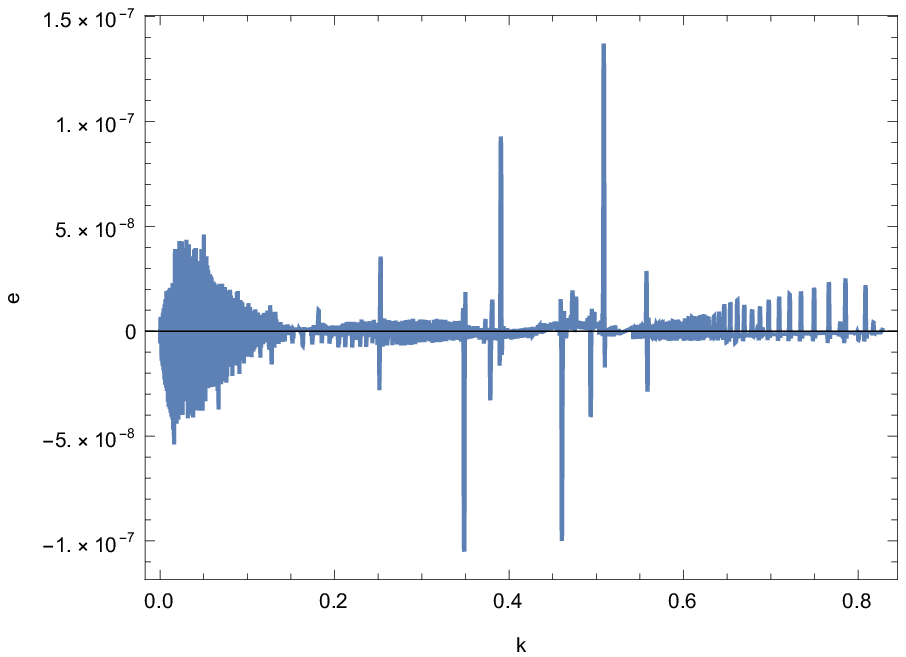}
\includegraphics[width=3in]{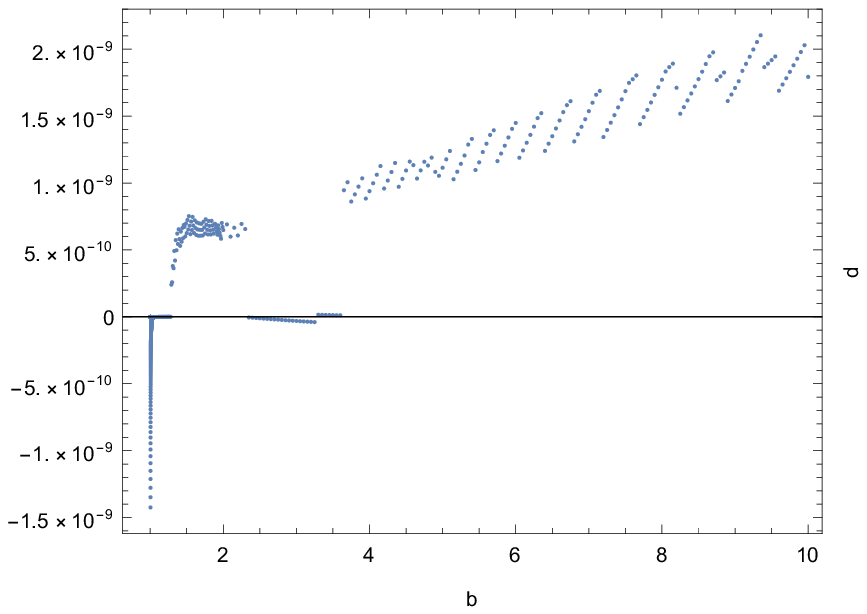}
\end{center}
  \caption{Numerical tests of the first law of thermodynamics for the
  perturbative conformal order (the left panel) and the conformal
  order at $K=0$ (the right panel) in $\mathfrak{M}_{PW}^b$ model.
} \label{error}
\end{figure}

In fig.~\ref{error} we test the first law of thermodynamics \eqref{firstlaw}
for the perturbative conformal order (see \eqref{pertconst}, the left panel)
and for the conformal order at $K=0$ (see \eqref{firstlaw2}, the right panel)
in $\mathfrak{M}_{PW}^b$ model.

\bibliographystyle{JHEP}
\bibliography{orderS3}

\end{document}